\documentstyle[prd,aps,multicol,epsf]{revtex}             %
\def \be{\begin{eqnarray}}
\def \ee{\end{eqnarray}}
\def \nn{\nonumber}

\def \De{\Delta}
\def \Th{\Theta} 
\def \epsi{\epsilon_1}
\def \epsii{\epsilon_2}
\def \udot{\dot{u}}
\def \cs{c_s^2}
\def \ca{c_a^2}
\def \caE{c_{aE}^2}
\def \tnabla{\tilde{\nabla}}
\def \veps {\varepsilon} 

\def \la{\langle}
\def \ra{\rangle} 
\def \qu{\quad} 
\def \dn{\left(1+w+ \frac 23 \frac{B^2}{\rho}\right)} 
\newcommand{\sfrac}[2]{{\textstyle{#1\over#2}}}
\newcommand{\hs}{\,-\,}
\newcommand{\al}[1]{\alpha_#1}
\newcommand{\bt}[1]{\beta_#1}
\newcommand{\gm}[1]{\gamma_#1}
\newcommand {\curly}[1]{\cal#1}
\begin{document}


\title{Dynamical Systems Approach to Magnetised Cosmological Perturbations}
\author{Stacey Hobbs\footnote{E-mail: stacey@maths.uct.ac.za} 
and Peter K.\ S.\ Dunsby\footnote{E-mail: peter@vishnu.mth.uct.ac.za}}
\address{Department of Mathematics and Applied Mathematics, University
  of Cape Town, Rondebosch 7701, Cape Town, South Africa.}
\date{\today}
\maketitle
\begin{abstract}
Assuming a large\hs scale homogeneous magnetic field, we follow the
covariant and gauge\hs invariant approach used by Tsagas and Barrow to
describe the evolution of density and magnetic field inhomogeneities
and curvature perturbations in a matter\hs radiation universe. We 
use a two parameter approximation scheme to linearize their 
exact non\hs linear general\hs relativistic equations for
magneto\hs hydrodynamic evolution. Using a two\hs fluid approach we 
set up the governing equations as a fourth order autonomous 
dynamical system. Analysis of the equilibrium points for the radiation
dominated era lead to solutions similar to the super\hs horizon 
modes found analytically by Tsagas and Maartens. We find that 
a study of the dynamical system in the dust\hs dominated era 
leads naturally to a magnetic critical length scale closely 
related to the {\it Jeans Length}. Depending on the size of 
wavelengths relative to this scale, these solutions show three 
distinct behaviours:  large\hs scale stable growing modes, 
intermediate decaying modes, and small\hs scale damped oscillatory 
solutions.
\end{abstract}


\vskip 4.5pc 

\begin{multicols}{2}
\section{Introduction}
Magnetic fields play an important role in our Universe. They appear
on all scales, from the solar system, through interstellar and
extra\hs galactic scales, to intra\hs cluster scales of several Mpc. Although 
magnetic field inhomogeneities have not yet been observed on scales 
as large as those exhibited by Cosmic Microwave Background
anisotropies, it is natural to expect that magnetic fields exits 
on such scales \cite{Bat}, and that they could play a role in the 
formation of large\hs scale structure. 

Magnetic fields in intergalactic space are notoriously invisible. In
the early 1950's the discovery of synchrotron radiation from the 
interstellar medium of the Milky Way led to the realization that 
it possessed a magnetic field. In 1957 Bolton and Wild \cite{bolt} 
suggested that the Zeeman splitting of a radio transition should 
be observable in the interstellar gas. This provided a direct way of 
measuring the strength of this uniform magnetic field. That almost ten 
years passed between this suggestion and the first detections
of this effect is an indication of the technical difficulty such
measurements posed at the time \cite{Kronberg}. Since then, advances 
in observational techniques have produced firmer estimates of magnetic 
field strengths in interstellar and intergalactic space. 
 
We now know that a large scale, organized magnetic field fills the
disk of the Milky Way. Studies of external galaxies indicate
that all disc galaxies are permeated by large scale magnetic 
fields \cite{Wielebinski}, and that {$\mu G$} level fields are common
in spiral galaxy discs and halos. Strengths of magnetic fields in 
intergalactic gas in `normal' galaxy clusters have been measured using 
Faraday Rotation Measures (RM) combined with X\hs ray data. This gives
typical field strengths for these fields of between $2$ and $6 \mu G$, 
\cite{KronNat}, which is comparable to the field strengths in the
denser interstellar medium in our galaxy. In 1993, Taylor and Perley 
\cite{taylor} discovered that these field strengths were exceeded by 
those in some less common, but more dense cooling flow clusters, where 
the field strength can be as high as $30 \mu G$. These fields show
ordered components on super\hs galactic scales \cite{KronNat}.

Much of our local universe consists of voids, regions containing very 
little baryonic matter. As yet, the weaker fields within these voids,
which may be relics of true cosmic primordial fields, remain unmeasured.
Primordial and protogalactic magnetic fields represent the large\hs
scale fields which could play a role in structure formation.
There are at present over thirty theories about the origin of cosmic 
magnetic fields at galactic and intergalactic scales. Battaner \&
Lesch \cite{Bat} look at astrophysical arguments to examine these
models. These can be divided into four main categories, based on when 
the fields are generated, namely: a) during inflation, b) in a phase 
transition after inflation, c) during the radiation dominated era, 
and d) after recombination. It is the large scale fields which were 
produced during inflation which are most likely to have implications
for structure formation. 

Investigating the effect of magnetic fields on structure formation is not
a recent endeavour. In 1971, Ruzmaikin \& Ruzmaikina \cite{RnR} gave a
Newtonian analysis of the growth of density perturbations in a
perfectly conducting medium with magnetic fields. Wasserman \cite{Was}
considered magnetic influence on galaxy formation, and angular momentum. 
Kim et al \cite{Kim} extend this by including the back\hs reaction of 
the fluid on to the field, and as a result, derived a {\it magnetic
Jeans length}. Battaner et al. \cite{BFJ} present relativistic analysis
of the evolution of magnetic fields and their influence on density 
inhomogeneities in a radiation dominated universe, while  Jedamzik et
al. and Subramanian \& Barrow \cite{JedSub} have considered magnetic 
dissipative effects at recombination. 

More recently Tsagas and Barrow \cite{TsaBar14,TsaBar15,Tsagas} 
have developed a covariant and gauge\hs invariant
approach to the analysis of magnetized density perturbations, in a 
universe model containing a perfect fluid. Using the covariant gauge\hs
invariant formalism of Ellis and Bruni \cite{ElBru}, they derived a
set of exact non\hs linear equations for general\hs relativistic
magneto\hs hydrodynamic evolution. On linearizing these equations 
around a flat Friedman\hs Robertson\hs Walker (FRW) model, they were 
able to identify the general relativistic corrections to earlier Newtonian
work, including the existence of a magneto\hs curvature coupling. 

Tsagas and Maartens \cite{TsaMar} extended these results, considering 
also shape distortion effects due to the fields as well as density 
and rotational perturbations They found explicit solutions to the 
perturbation equations for the radiation and dust eras, as well as 
pure\hs magnetic density perturbations. They also identified and 
analysed other sources of the magnetic effects. 

When studying the evolution of density perturbations, the dynamics 
of the perturbed universe can be represented by an autonomous 
dynamical system, described by a set of coupled differential equations 
for the gauge\hs invariant perturbation variables \cite{Wasz,Bruni,BruPio}. 
This allows the stability behaviour of the model to be investigated 
with relative ease, with no need to solve for the background
variables. Furthermore, stationary points of the dynamical 
system correspond to exact or approximate analytic solutions of 
the linearised perturbation equations, thus providing a useful tool
for obtaining solutions in cosmologically interesting situations. 

This paper is outlined as follows. In section \ref{Equations}, we set 
up the evolution equations of the density and magnetic field
inhomogeneities, and the curvature perturbations. We discuss in detail 
the approximations and linearisation procedure in sections 
\ref{Approximations} \& \ref{Linearization} and identify an extra term 
in the linearized propagation equation for the spatial curvature
${\cal K}$. In section \ref{DS}, we set up a five dimensional
autonomous dynamical system equivalent to the linearized propagation 
equations derived in section \ref{Equations} \hs these equations give
a detailed description of the full phase\hs space of solutions for 
perturbation dynamics in a magnetized dust\hs radiation universe. 
The analysis and discussion of this system, with particular emphasis
on the three dimensional invariant sets representing the initial
radiation and final dust dominated states are given in section 
\ref{Analysis}. Finally, conclusions are given in section \ref{Summary}.
\section{Defining Variables}
\subsection{Spacetime Splitting}
For all of the following equations, we will use standard units defined 
by $8\pi G=c=1$. In cosmology, the average velocity of matter at each 
spacetime event defines a unique 4\hs velocity vector: 
$u^a = \frac{dx^a}{d \tau}$ with $u_a u^a = -1$.  
The {\it fundamental fluid\hs flow lines} defined by this 
vector field are a congruence of worldlines, carrying 
the {\it fundamental observers}. The projection tensor $h_{ab} =
g_{ab} + u_a u_b$, (with $h^a{}_c h^c{}_b = h^a{}_b$, and $h^a{}_a=3$)
projects into the local rest spaces of comoving observers.

The covariant derivative of any tensor $T^{ab}{}_{cd}$ may be split 
into a time derivative along the fluid flow, 
\begin{equation}
\dot{T}^{ab}{}_{cd} = u^e \nabla_e T^{ab}{}_{cd}\;,
\end{equation}
and a totally projected covariant derivative orthogonal to the fluid
flow \cite{Carg} 
\begin{equation}
\tnabla_e T^{ab}{}_{cd} = h^a{}_f h^b{}_g h^p{}_c h^q{}_d h^r{}_e 
\nabla_r T^{fg}{}_{pq}\;,
\end{equation}
with total projection on all free indices. It is worth noting that if 
$u^a$ has non\hs zero vorticity, $\tnabla$ is not a proper 3\hs 
dimensional covariant derivative. Angle brackets denote orthogonal 
projection of vectors $V^{\la a \ra}= h^a{}_b V^b$,
and the projected symmetric trace\hs free (PSTF) part of tensors 
$T^{\la ab \ra} = [{h^{(a}}{}_c h^{b)}{}_d - \frac{1}{3} h^{ab} 
h_{cd}]T^{cd}$ \cite{maartens}. 

Following \cite{TsaMar}, we can define a covariant {\it spatial divergence} 
and {\it curl} that generalize the Newtonian operators to curved spacetime:
\begin{eqnarray}
{\rm div} V = \tnabla^a V_a &,& \quad ({\rm div } T)_a = \tnabla^b
T_{ab}\;, \nonumber \\
{\rm curl} V_a = \varepsilon_{abc} \tnabla^b V^c &,& \quad
{\rm curl} T_{ab} = \varepsilon_{cd(a} \tnabla^c {T_{b)}}^d\;,
\end{eqnarray}
where $\varepsilon_{abc} \equiv \eta_{abcd} u^d$ is the projected covariant
permutation tensor.
\subsection{Kinematic Quantities}
The projection tensor, together with the velocity 4\hs vector, split 
the covariant derivative of $u_a$ into irreducible basic kinematic 
quantities \cite{TsaBar14}
\cite{Carg}:
\be
\nabla_b u_a &=& -u_b \udot_a + \tnabla_b u_a \nn \\
&=& -u_b \udot_a + \sfrac13 \Theta h_{ab} + \sigma_{ab} + \omega_{ab}\;,
\label{eq:delu}
\ee
where $\udot_b = \nabla_a u_b u^a$, $\Theta = \tnabla_a u^a$,
$\sigma_{ab} = \tnabla_{\la a}u_{b \ra}$ and 
$\omega_{ab} = \tnabla_{[b} u_{a]}$.  
These are, respectively, the 4\hs acceleration, the expansion scalar, 
the shear tensor and the vorticity tensor. The expansion scalar is used to 
introduce a representative length scale $a$ along the observers' worldline
which in turn defines the Hubble expansion rate, $\frac{\dot{a}}{a} = 
\frac{1}{3}\Th = H$. The evolution of the expansion scalar is described
by Raychaudhuri's equation:
\begin{equation}
\dot{\Th} - A + 2(\sigma^2 - \omega^2) + \sfrac13 \Th^2 
+ \sfrac12 (\rho + 3p + B^2) = 0\;,
\label{eq:Ray}
\end{equation}
where $A = \nabla^a \udot_a$, $\sigma^2 = \case{1}/{2}\sigma^{ab}
\sigma_{ab}$ and $\omega^2 = \case{1}/{2}\omega^{ab} \omega_{ab}$ 
(see \cite{Carg} for details).

Local curvature is described by the Ricci tensor $R_{ab}$ while non\hs 
local tidal forces and gravitational radiation are described by the 
'electric', $E_{ab}$ and 'magnetic', $H_{ab}$ parts of the Weyl 
tensor. These are a pair of symmetric, traceless and completely 
spacelike second order tensors, defined by:
\begin{equation}
 E_{ab} = E_{\la ab \ra} =  C_{acbd} u^c u^d\;,
\end{equation}
\begin{equation}
 H_{ab} = H_{\la ab \ra} =  \sfrac12 \veps_{acd} {C^{cd}}_{be} u^e\;.
\end{equation}
\subsection{The Electromagnetic Field} 
The electromagnetic field is represented by the Maxwell tensor 
$F_{ab} = F_{[ab]}$. As seen by an observer with 4\hs velocity 
$u_a$, the field tensor splits into an electric $E_a$ and a 
magnetic $B_a$ part (see \cite{Carg} for details):
\begin{equation}
E_a = F_{ab}u^b = -F_{ab}u^a\;,
\end{equation} 
\begin{equation}
B_a = \sfrac12 \veps_{acd} F^{cd}\;,
\end{equation}
where $E_a u^a = 0$, and $B_a u^a$ = 0. 
\section{Equations} \label{Equations}
\subsection{Maxwell's Equations}    
Using the above decomposition, Maxwell's equations can be 
written as follows:
\begin{equation}
\nabla_b  F^{ab} = J^a\;,
\label{eq:4Max1}
\end{equation}
and 
\begin{equation}
\nabla_{[c} F_{ab]} = 0\;,
\label{eq:4Max2}
\end{equation}
where $J^a$ is the 4\hs current which generates the electromagnetic
field and obeys the conservation law $\nabla_a J^a = 0$.

These equations can be split into spatial and temporal parts using 
the notation defined above:
\begin{equation}
{\rm div}E + 2 \omega^a B_a = q\;,
\label{eq:Max1}
\end{equation}

\be
\sigma^{ab} E_b &+& \veps^{abc} \omega_c E_b - \frac 23 \Theta E^a
+ \veps^{abc} \udot_b B_c \nn \\
&+& {\rm curl} B^a = {\dot{E}}^{<a>} + j^a\;,
\label{eq:Max2}
\ee

\begin{equation}
{\rm div}B = 2 \omega^a E_a\;,
\label{eq:Max3}
\end{equation}
and 
\be
\sigma^{ab} B_b &-& \veps^{abc} \omega_b B_c - \frac 23 \Theta B^a -
\veps^{abc} \udot_b E_c \nn \\
&-& {\rm curl}E^a = {\dot{B}}^{<a>}\;.
\label{eq:Max4}
\ee
The 4\hs current $J^a$ enters into equations (\ref{eq:Max1}) 
and (\ref{eq:Max2}) through the charge density $q = -J^a u_a$ and 
the projected current $j^a = J^{<a>}$. The vorticity vector
$\omega_a$ is  related to the vorticity tensor via $\omega_{ab} 
\equiv \veps_{abc} \omega^c$.
\subsection{Gauge\hs invariant Variables}
Following \cite{TsaMar} and \cite{TsaBar14}, the key covariant and
gauge\hs invariant variables describing inhomogeneity in the fluid and 
the magnetic field are the comoving spatial gradients of the 
{\it energy density} $\rho$, the {\it expansion} $\Theta$ and the 
{\it field density} $B^2$:
\begin{equation}
\Delta_a ={\frac{a \tnabla_a \rho}{\rho}},   
\qu  {\Theta}_a = a \tnabla_a \Theta, 
\qu     {\curly B}_a = \tnabla_a B^2\;,
\label{eq:defvar}
\end{equation} 
where $B^2 = B_a B^a$. Other basic variables which appear in the exact
equations are the spatial gradient of the pressure, $Y_a = \tnabla_a
p$, the anisotropic pressure generated by the magnetic field, 
$\Pi_{ab} = \frac{1}{3} B^2 h_{ab} - B_a B_b$, and the
comoving spatial gradient of the field vector: $B_{ab} = a \tnabla_b B_a$.
\subsection{Medium and Propagation Formulae}
The matter description under consideration is a mixture of dust 
and radiation which share the same 4\hs velocity $u^a$, 
interacting only through gravity. In general, the 
evolution equations for density perturbations couple to an 
entropy evolution equation through the equation of state $p = p(\rho,
s)$, where $s$ is the entropy density. However, since entropy 
perturbations are only important on very small scales, we will assume 
adiabatic perturbations at all times, thus ignoring any entropy 
contributions \cite{DBE}.

The total fluid equations take the same form as those for a perfect
fluid of infinite conductivity. The infinite conductivity
approximation allows the electric field to be omitted 
from Maxwell's Equations while spatial currents are preserved 
\cite{TsaBar14}. 

In this case, Maxwell's equations (\ref{eq:Max1}) - (\ref{eq:Max4}) generate
three constraints \cite{TsaMar},
\begin{equation} 
\omega^a B_a = \sfrac12 q\;,
\end{equation}

\begin{equation}
{\rm curl} B^a = \veps^{abc} B_b \udot_c + j^a\;,
\label{eq:curlB}
\end{equation}

\begin{equation}
{\rm div}B = 0\;,
\label{eq:divB}
\end{equation}
and one propagation equation
\begin{equation} 
{\dot{B}}^{<a>} = \sigma^{ab} B_b + \veps^{abc} B_b \omega_c
- \sfrac23 \Theta B^a\;.
\label{eq:dotB}  
\end{equation}
\subsection{Conservation Equations}
Since we are only considering adiabatic perturbations, we only need 
the total fluid equations. The energy and momentum density
conservation equations are respectively \cite{TsaBar14}:
\begin{equation}
\dot{\rho} + \rho(1 + w) \Theta = 0\;,
\label{eq:rhodot}
\end{equation}
and
\begin{equation}
\rho(1 + w + \frac{2 B^2}{3 \rho}) \udot_a + Y_a -  \frac 2a B_{[ab]} B^b
+ \udot^b \Pi_{ba} = 0\;,
\label{eq:udot}
\end{equation}
where $w=\frac{p}{\rho}$.  Ignoring entropy perturbations, 
the pressure can be written as a function of the energy density 
only, i.e. $p = p(\rho)$.  It follows that the relationship between 
$Y_a$ and $\De_a$ is given by
\begin{equation}
a Y_a = \rho \cs \De_a\;,
\label{eq:Ya}
\end{equation}
where $\cs = \frac{dp}{d \rho}=\frac{\dot{p}}{\dot{\mu}}$ is the speed
of sound in the fluid.
\subsection{Exact Propagation Equations for the Inhomogeneity Variables}
The propagation equations for the various spatial gradients are given in
\cite{TsaBar14}. They are, firstly, the equation for the 
{\it comoving fractional spatial gradient of the energy\hs density}: 
\begin{eqnarray}
{\dot{\De}}_{<a>} &=& w \Th \De_a - ({\sigma^b}_a + {\omega^b}_a)\De_b
- \left( {1 + w} \right) {\Th}_a \nn \\
&-& \frac {2 \Theta}{\rho} {B}_{[ab]} B^b
+ \frac{2 a \Th B^2}{3 \rho} \udot_a
+ \frac{a \Theta}{\rho} \udot^b \Pi_{ba}\;,
\label{eq:Dprop}
\end{eqnarray}
secondly, the equation that governs {\it the spatial
gradient of the expansion}:
\begin{eqnarray}
{\dot{\Th}}_{<a>} &=& - \frac{2 \Theta}{3}  
{\Th}_a - ({\sigma^b}_a + {\omega^b}_a){\Th}_b 
- \sfrac12\rho \De_a 
+ B_{ba}B^b  \nn \\ 
&-& 3 B_{[ab]}B^b
+ a{\curly R} \udot_a
+ \sfrac32 a \udot^b \Pi_{ba} + a A_a \nn \\
&-& 2 a \tnabla_a(\sigma^2 - \omega^2)\;,
\label{eq:Thprop}
\end{eqnarray}
where $A_a = {h_a}^b A_{;b} = \tnabla_a A$, $A=\tnabla^a \udot_a$ and 
\begin{equation}
{\curly R} = \case{1}/{2}K + A - 3(\sigma^2 - \omega^2)\;,
\end{equation}
with
\begin{equation}
K=\rho + \sfrac12 B^2 - \sfrac13 \Theta + \sigma^2 - \omega^2 
\end{equation}
representing the 3\hs Ricci scalar of the observers instantaneous rest space. 

Finally, the equation describing the evolution of the {\it orthogonal spatial 
gradient of the magnetic field} is given by:
\begin{eqnarray}
a^{-2}{h_a}^e{h_b}^p (a^2 {B}_{ep}) \dot{} &=&
-{B}_{ae}({\sigma^e}_b + {\omega^e}_b)
+({\sigma_a}^e + {\omega_a}^e){B}_{eb} \nn \\
&-& \sfrac 23 B_a{\Th}_b + 2a B_e{\sigma^e}_{[a} \udot_{b]} \nn \\
&-& 2aB_e {\omega^e}_{(a} \udot_{b)}
+ a B^e \tnabla_b (\sigma_{ae} + \omega_{ae}) \nonumber \\
&-& \sfrac13 \Theta a (2B_a \udot_b + \udot_a B_b) \nn \\
&+& a \udot^e B_e (\sigma_{ab} + \omega_{ab} + \sfrac13 \Theta h_{ab}) 
\nonumber \\
&-& a {h_a}^e R_{epbq} B^p u ^q\;.
\label{eq:Bprop}
\end{eqnarray}
\section{Detailed Discussion of Approximations}
\label{Approximations}
In the papers by Tsagas, Maartens and Barrow \cite{TsaMar} \hs
\cite{Tsagas}, the cosmic magnetic field is treated as a coherent test 
field propagating on the background. The large\hs scale 
magnetic field $B_a$ is assumed to be too weak to destroy the spatial 
isotropy of the background spacetime, which is taken to be a flat 
Friedman\hs Robertson\hs Walker (FRW) model\footnote{Spatial flatness 
is necessary for the 3\hs Ricci scalar to be gauge\hs
invariant.}. This is an acceptable physical approximation when
the magnetic field density is small compared to the energy density
of the fluid i.e.  $B^2 / \rho << 1$.  This issue is discussed in \cite{Zel},
where Zel'dovich calculates that $B^2/ \rho < 8\times 10^{-5}$ if 
the model is to be applicable during the whole of the radiation dominated era. 
Furthermore, the isotropy of the microwave background can be used
to place strong limits on the magnitude of the magnetic field 
\cite{adams,Barrow,BFS}.

Other authors have used alternative approaches to the problem 
of maintaining isotropy in the background in the presence of 
magnetic fields. Battaner et al. \cite{Bat} show that a mean 
magnetic field is incompatible with a Robertson\hs Walker metric 
and therefore assume that there is {\it no} mean magnetic field 
on cosmological scales, i.e. $<B_a> = 0$. However, they include 
the presence of magnetic fields in smaller cells, with field
directions random on larger scales. Thus, although there is no mean 
magnetic field, the model includes an average magnetic energy 
density $<B^2> \neq 0$. Kim et al. \cite{Kim} also follow this
approach, considering that, at recombination, field directions are
randomly oriented on scales smaller than the Hubble radius.

In order to maintain the coupling between magnetic irregularities 
and energy density perturbations in a straightforward way, a perfectly 
conducting medium is introduced.  Looking at the covariant form of 
Ohm's law \cite{TsaBar14}, we have,
\begin{equation}
J^a + J_b u^b u^a = \sigma E^a\;,
\end{equation}
where $\sigma$ represents the conductivity of the medium.  
Projecting into the rest space of a fundamental observer yields
\begin{equation}
J^{<a>} = \sigma E^a\;.
\end{equation}
To have a vanishing electric field, while maintaining non\hs zero 
spatial currents ($J^{<a>} \neq 0$), the conductivity of the medium 
must be infinite, i.e. $\sigma \rightarrow \infty$.

An alternative approach would be to assume a pure magnetic field 
with no electric field and no spatial currents. This would 
reduce (\ref{eq:Max2}) to
\begin{equation}
\veps^{abc} \udot_b B_c  + {\rm curl} B^a = 0\;,
\end{equation}
which is equivalent to
\begin{equation}
  \udot_{[b} B_{c]} + \sfrac1a B_{[cb]} = 0\;.
\end{equation}
Although this appears to couple the acceleration to the field
gradient, when this relation is inserted into (\ref{eq:udot}), 
all the magnetic terms cancel out. This effectively decouples 
the magnetic field from energy density inhomogeneities.
\section{Approximation Scheme} \label{Linearization}
In order to linearize the equations (\ref{eq:Dprop}), (\ref{eq:Thprop}) \&
(\ref{eq:Bprop}), we introduce {\it two} smallness parameters.
The first, {$\epsi$} is used to measure the extent to which  
the gauge\hs invariant variables deviate from zero (their value in a 
flat FRW universe). The other parameter $\epsii$ is a measure of the 
Alfv\'{e}n speed  $B^2 / {\rho}$. 
\subsection{Zero\hs  order quantities} 
The energy density $\rho$, pressure $p$ and expansion $\Th$ 
do not vanish in the background spacetime. The magnetic field, 
$B_a$, is treated as a small test field propagating 
on the background. It follows that these variables can 
regarded as zeroth order in our approximation scheme.  
\subsection{First\hs  order quantities [in $\epsi$]}
In order for the metric in some region of spacetime $\cal{U}$ to 
be written in a perturbed FRW form, the following inequalities 
must hold for the smallness parameter $\epsi$ \cite{Tsagas,SME}:
\begin{eqnarray}
\frac{\sigma}{H} < \epsi, \qu \frac{\omega}{H} < \epsi, \qu
\frac {|E_{ab}|}{H^2} < \epsi, \qu  \frac{|H_{ab}|}{H^2} < \epsi, 
\nonumber \\
\frac{|\tnabla_a \rho|}{\rho H} < \epsi, \qu
\frac{|\tnabla_a \Theta|}{H^2} < \epsi.
\end{eqnarray}
where
$|E_{ab}| \equiv (E_{ab}E^{ab})^{1/2}, |\tnabla_a \rho| \equiv 
(\tnabla_a \rho \tnabla^a \rho)^{1/2}$ etc. 

Tsagas \cite{Tsagas} extends this definition to a magnetized universe
by arguing that closeness to a spatially flat FRW spacetime is
maintained when additional restrictions are imposed as follows:
\begin{eqnarray}
\frac{|\Pi_{ab}|}{H^2} < \epsi, \qu \frac{|\tnabla_b B_a|}{H|B_a|} < \epsi
\quad {\rm and} \quad \frac{|K|}{H^2} < \epsi.
\end{eqnarray}
In an exact flat FRW spacetime, all quantities of order $\epsi$
vanish identically. Thus, $\sigma_{ab}$ , $\omega_{ab}$,
$\udot_{a}$, \{$A$, $A_{a}$\}, \{$E_{ab}$, $H_{ab}$\}, \{$\De_a$,
${\Theta}_a$\}, \{$B_{ab}$, ${\curly B}_{a}$\}, \{${\curly K}_a = a^2 K$\} 
are all considered to be first order in $\epsi$. 

Note that although the magnetic field vector $B_a$ is considered to
be a zeroth order quantity, its magnitude must remain small so that it
does not disturb the isotropy of the background. This ensures that the 
the anisotropic pressure generated by the magnetic field, 
$\Pi_{ab}$, is negligible in the background, and therefore 
$\Pi_{ab}$ may also be regarded as first order in $\epsi$. 

Linearization of the above equatioons is implemented as follows: 
All terms higher than 
first order in $\epsi$ are dropped, as well as terms higher than 
first order in the Alfv\'{e}n parameter $\epsii$. Terms like 
$\epsi \epsii$ are however kept.

At the end of the calculation, terms first order in $\epsii$ are
dropped relative to zero order terms in the coefficients 
of quantities that are first order in $\epsi$. This is 
permissible since the magnetic field is very weak ($ B^2 / {\rho} <<
1$). However, this must be done last, otherwise terms may be 
dropped relative to others which could later vanish.
\section{Linear Equations}
\subsection{Conservation Equations}
Linearization leaves equation (\ref{eq:rhodot}) unchanged, but 
slightly modifies the momentum density equation (\ref{eq:udot}):
\begin{equation}
\rho(1+w+  \frac{2 B^2}{3 \rho}) \udot_a + Y_a 
- \frac{2}{a} B_{[ab]} B^b = 0\;.
\label{eq:udotl}
\end{equation}
This, together with (\ref{eq:Ya}), gives a useful expression for the 
acceleration vector:
\begin{equation}
\udot_a  = \frac{1}{\dn a} \left( \frac{2}{\rho} B_{[ab]} B^B
- \cs \De_a \right)\;.
\label{eq:a}
\end{equation}
It follows that the divergence of the acceleration $A$ 
is given by \footnote{In deriving this, $\tnabla_a w$ and $\tnabla_a \cs$ are 
treated as first order in $\epsi$,  
as in \cite{TsaBar15}}
\begin{equation}
A = \frac{1}{\dn} \left(-\frac{\cs}{a^2} \De +  \frac{B^2}{3 \rho} K
- \frac{B^2}{2 \rho a^2} {\curly B} \right)\;.
\label{eq:A}
\end{equation}
\subsection{Linearized Propagation Equations} 
The linearized propagation equations for the 
key gauge\hs invariant quantities are obtained by dropping terms 
of ${\cal O}(2)$ or higher in both linearisation parameters. Note 
that terms of mixed order ($\epsi \epsii$) and  first order 
in $\epsii$ are kept to start with (but may be dropped later).

In these equations, the expansion scalar $\Th$ is replaced by 
the equivalent expression involving the Hubble parameter i.e. $\Th =
3H$. Also, to first order, all projected time derivatives of 
quantities that are first order in $\epsi$ are equal to their normal 
time derivatives (e.g. ${\dot{\De}}_{<a>} = {\dot{\De}}_a$).

It follows that the energy density propagation equation
(\ref{eq:Dprop}) becomes:
\begin{eqnarray}
{\dot{\Delta}}_a &=& 3wH \Delta_a - (1+w) \Theta_a \nn \\ 
&-& \frac{6H}{\rho} B_{[ab]}B^b
+ \frac{2a H B^2}{\rho} \udot_a\;.
\label{eq:Deltaa}
\end{eqnarray}
The equation for the expansion gradient (\ref{eq:Thprop}) involves the
spatial gradient of the divergence of the acceleration $A_a$. This can be
determined using expression (\ref{eq:A}), the identities for commutations
between spatial gradients and time derivatives given in Appendix
\ref{ids}, and the spatial gradient of the 3\hs curvature $K$, given by 
\begin{equation}
a\tnabla_a K =  2\rho \De_a + a{\curly B}_a - 4H \Th_a\;.
\label{eq:Kgrad}
\end{equation}
Using these expressions, the evolution of the expansion gradient 
(to first order in $\epsi$) is 
\begin{eqnarray}
{\dot{\Th}}_a &=& -2H \Th_a - \sfrac12 \rho \De_a - \sfrac12 a {\curly B}_a
-3 B_{[ab]} B^b \nonumber \\
&-& \frac{\cs}{\dn} \tnabla^2 \De_a 
-\frac{a}{2 \rho \dn} \tnabla^2 {\curly B}_a \nonumber \\
&-& \left [ \frac{6 \cs(1+w)}{\dn}  + \frac{4 B^2}{\rho \dn} \right]
aH \tnabla^b \omega_{ab} \nonumber \\
&+& \frac{2B^2}{3\rho \dn} \rho \De_a
+ \frac{B^2}{3 \rho \dn} a {\curly B}_a \nonumber \\
&-& \frac{4B^2}{3\rho \dn} H {\Th_a}\;.
\label{eq:Thetaa}
\end{eqnarray}
\subsection{Scalar Equations}
Focusing on the growth, or decay, of density inhomogeneities, the vector field
$\De_a$ contains more information than is necessary. We can extract the
required information by considering a local decomposition of the 
spatial gradient of $\De_a$ first introduced in
\cite{EBH}\footnote{This decomposition in analogous to the 
  decomposition of the first covariant derivative of the 4\hs
  velocity: (\ref{eq:delu}).}:
\begin{equation}
\De_{ab} \equiv a \tnabla_b \De_a = W_{ab} + \Sigma_{ab}
+ \sfrac13 \De h_{ab}\;,
\label{eq:decomp}
\end{equation}
where $W_{ab} \equiv \De_{[ab]}$ represents rotations of the density
gradient $\De_a$, $\Sigma_{ab} \equiv \De_{(ab)} - \De h_{ab}$
describes the variations of $\De_a$ associated with pancake\hs  
or cigar\hs  like structures, and finally
$\De \equiv \De^a{}_a = a \tnabla^a\De_a$ is related to spherically 
symmetric gravitational clumping of matter. It is this scalar variable 
that is important when examining structure formation.

We also need to consider the following complementary scalar 
variables \cite{TsaBar15}:
\begin{eqnarray}
{\curly Z} \equiv a \tnabla^a \Th_a\;, \qu
{\curly B} \equiv \frac{a^2}{B^2} \tnabla^2 B^2\;, \qu
{\curly K} = a^2 K\;,
\label{defs}
\end{eqnarray}
which represent spatial divergences in the expansion gradient, the
energy density gradient of the magnetic field, and perturbations in
the spatial curvature respectively. 

As before, we linearize the scalar propagation equations by dropping terms of
order $\epsi^2$ or $\epsii^2$, but retain all terms that are first
order in $\epsii$.

Equations describing the propagation of these scalars are:
\begin{eqnarray}
\dot{\Delta} &=& 3wH{\Delta} - (1+w) {\curly Z} +
{\frac{3 H  B^2} {2 \rho }} {\curly B}\nonumber\\
&-&{\frac{H B^2}{\rho}} {\curly K}
-{\frac{2 \cs B^2}{\rho \dn}} H {\Delta}\;,
\label{eq:deldot}
\end{eqnarray} 
\begin{eqnarray}
\dot{\curly Z} = &-& 2H {\curly Z} - \frac{\rho}{2} \Delta
- \frac{B^2}{2} {\curly K}
+ \frac{B^2}{4} {\curly B}
- \frac{\cs}{\dn} \tnabla^2 \Delta \nonumber \\
&-& \frac{B^2}{2 \rho \dn} \tnabla ^2 {\curly B}
+ \frac {2 B^2}{3 \rho\dn} \rho \Delta \nonumber \\
&+& \frac {B^2}{3 \rho \dn} B^2 {\curly B}
- \frac {4}{3}
\frac{B^2}{\rho \dn} H {\curly Z}\;, \nonumber \\
\label{eq:Zdot}
\end{eqnarray}
\begin{eqnarray}
\dot{\curly B} = &-& \frac{4}{3} {\curly Z} +
\frac{4 H \cs}{\dn} \Delta
- \frac{4 H B^2}{3 \rho \dn}{\curly K}
\nonumber \\
&+& \frac{2 H B^2}{\rho\dn}{\curly B}\;,
\label{eq:Bdot}
\end{eqnarray}
\begin{eqnarray}
\dot{\curly K} = & & \frac{4 H \cs}{\dn} \Delta
+ \frac{2 H B^2}{\rho\dn} {\curly B}
\nonumber \\
&-& \frac {4 H B^2}{3 \rho\dn}{\curly K}\;.
\label{eq:Kdot}
\end{eqnarray}
\subsection{Final Linearized System}
Combining equations (\ref{eq:deldot}\hs \ref{eq:Kdot}), and finally 
dropping terms of order $\epsii$ with 
respect to zero order quantities, we obtain a second\hs order 
differential equation for the scalar energy density perturbations:
\begin{eqnarray}
\ddot{\De} &=& - (2+ 3\cs - 6w)H \dot{\De} 
+ \sfrac12 (1 - 6 \cs + 8w - 3w^2) \rho \De\nonumber \\
&+& \cs \tnabla^2 \De  
- \frac{2 B^2 H}{\rho (1 +w)} \dot{c}_s^2 \De 
- \sfrac12 (1 - 3\cs + 2w) \rho \ca {\curly B} \nonumber \\
&+& \sfrac12 \ca \tnabla^2 {\curly B} 
+ \sfrac13 (2 - 3\cs + 3w) \rho \ca {\curly K}\;. 
\end{eqnarray}
It is not immediately obvious that the term containing $\dot{c}_s^2$
is negligible with respect to the other terms involving $\De$. 
In order to clarify this, we can write $\dot{\cs}$ in terms
of $w$ by noting
\begin{eqnarray}
\dot{c}_s^2 &=& \frac{d(\cs)}{dw} \dot{w} \nonumber \\
&=& \frac{4 (w - \cs)}{(1+w)} H\;.
\label{eq:csdot}
\end{eqnarray}
In what follows we will consider the above equations for values 
of $w \in [0, \sfrac13]$, but more specifically, we will 
look at the dust  $w=0$ and radiation dominated $w = \sfrac13$ eras. 
From equation (\ref{eq:wprop}) we see that $w = 0 \Rightarrow \cs = 0$ 
and that $w = \sfrac13 \Rightarrow \cs = \sfrac13$. Thus, the term 
drops away during both these eras and is negligible at all other times.

Without that term, the second order propagation equation 
for $\De$ becomes
\begin{eqnarray}
\ddot{\De} &=& - (2+ 3\cs - 6w)H \dot{\De} 
+ \sfrac12 (1 - 6 \cs + 8w - 3w^2) \rho \De \nonumber \\
&+& \cs \tnabla^2 \De  
- \sfrac12 (1 - 3\cs + 2w) \rho \ca {\curly B} 
+ \sfrac12 \ca \tnabla^2 {\curly B} \nonumber \\
&+& \sfrac13 (2 - 3\cs + 3w) \rho \ca {\curly K}\;. 
\label{eq:delfinal}
\end{eqnarray}
Finally, the completely linearized propagation equations 
for ${\curly B}$ and ${\curly K}$ are
\begin{equation}
\dot{\curly B} = \frac{4}{3(1+w)} \dot{\De} + 
4H \frac{(\cs - w)}{(1+w)} \De\;,
\label{eq:Bfinal}
\end{equation}
and
\begin{equation}
\dot{\curly K} = \frac{4 H \cs}{(1+w)} \De
- \frac43 \frac{B^2 H}{\rho (1+w)} {\curly K}
+ \frac{2 B^2 H}{\rho (1+w)} {\curly B}\;.
\label{eq:Kfinal}
\end{equation}
Equations (\ref{eq:delfinal}) and (\ref{eq:Bfinal}) are identical 
to those in \cite{TsaMar}, but as a result of slightly different 
linearisation \footnote{At this point Tsagas and Maartens \cite{TsaMar}
expand the LHS of equation (\ref{eq:Kfinal}) using the definition of 
${\curly K}$ (see equation (\ref{defs})): $\dot{\curly K}=2H{\curly K}
+a^2\dot{K}$ and then drop the second term on the RHS of (\ref{eq:Kfinal})
relative to $2H{\curly K}$.}, equation (\ref{eq:Kfinal}) has an 
extra term compared to the corresponding result in \cite{TsaMar}. 
\section{Dynamical Systems Analysis}
\label{DS}
Instead of attempting to find exact solutions to equations
(\ref{eq:delfinal}\hs \ref{eq:Kfinal})
for a dust\hs radiation background, we instead follow Bruni and Piotrowska
\cite{BruPio} and perform a qualitative analysis of the above 
system of differential equations. 
\subsection{The System}
It is useful to change the independent variable from 
proper time to a function of the scale factor $a$: with the
choice $\tau = \ln a$ (which yields $d\tau / dt = H$), and using the
standard harmonic decomposition described in Appendix \ref{sec:appendB},
equations (\ref{eq:delfinal}), (\ref{eq:Bfinal}) and (\ref{eq:Kfinal}) become, 
\begin{eqnarray}
\De_{(k)} '' &=& - \sfrac12 (1 + 6\cs - 15 w) \De_{(k)} ' 
+ \sfrac32 ( 1- 6\cs + 8w - 3w^2) \De_{(k)} \nonumber \\
&-& \cs \frac{k^2}{a^2 H^2} \De_{(k)}  
- \sfrac32 ( 1- 3\cs + 2w) \ca {\curly B}_{(k)} 
- \frac{\ca}{2} \frac{k^2}{a^2 H^2} {\curly B}_{(k)} \nonumber\\
&+& (2 - 3\cs + 3w) \ca {\curly K}_{(k)}\;,
\label{eq:D1}
\end{eqnarray}
\be
{\curly B}_{(k)}' = \frac{4}{3(1+w)} \De_{(k)} ' + \frac{4( \cs - w)}{(1+w)} \De_{(k)}\;,
\label{eq:B1}
\ee
and
\be
{\curly K}_{(k)} ' = \frac{4\cs}{(1+w)} \De_{(k)} - \frac{4 \ca}{3 (1+w)} {\curly K}_{(k)}
+ \frac{2 \ca}{(1+w)} {\curly B}_{(k)}\;,
\label{eq:K1}
\ee
where a prime denotes differentiation with respect to $\tau$. From now 
on we will drop the subscript $(k)$.
\subsection{Coefficients in Terms of $w$}
In order to close the system, propagation equations are 
needed for all background variables. It turns out however 
that the coefficients of the above differential equations can 
all be written explicitly as functions of the equation of state 
parameter $w$, thus only an evolution equation for $w$ needs to 
be found. To achieve this, we normalise the scale factor $a$
at dust\hs radiation equi\hs density by introducing the variable
$S=a/a_E$ (see Ehlers \& Rindler \cite{Ehlers-Rindler} and 
Padmanabhan \cite{paddy}).

Since the two fluids are coupled only through gravity, the energy 
conservation equation is obeyed separately for each component. This 
gives $\rho_d= \sfrac{\rho_E}{2} S^{-3}$ and $\rho_r =
\sfrac{\rho_E}{2} S^{-4}$ for dust and radiation respectively, where 
$\rho_E$ is the total energy density at equi\hs density. Total energy 
conservation yields $\rho = \sfrac{\rho_E}{2}(S^{-3} + S^{-4})$. The 
only contribution to the total pressure comes from radiation
component: $p = p_r = \sfrac{\rho_r}{3} = \sfrac{\rho_E}{6} S^{-4}$. 
It is now straightforward to find $w$ in terms of $S$ \cite{BruPio}:
\be
w = \frac{1}{3(S+1)}\;,
\label{eq:woS}
\ee
which can be inverted to give 
\be
S= \frac{1 - 3w}{3w}\;.
\label{eq:Sow}
\ee
In this way the expansion of the universe model is parameterised 
by $w$, with $w \in [0, \sfrac13]$ and varies from a pure 
radiation\hs dominated ($t\rightarrow 0$) to a pure dust\hs 
dominated ($t\rightarrow\infty$) phase.

These results (\ref{eq:woS}) and (\ref{eq:Sow}) allow us to write  
the coefficients that occur in the perturbation equations 
as functions of $w$ only (see Appendix \ref{sec:appendC}).

The energy density of the magnetic field $B^2$ has a radiation\hs like
propagation equation $B^2 = {B_E}^2 S^{-4}$, hence the Alfv{\'{e}}n 
speed can be written as
\be
\ca = 6w\caE\;.
\ee
The propagation equation for $w$ can now be written down. It is
\be
w' = 3w(w- \sfrac13)\;.
\label{eq:wprop}
\ee
In terms of $w$, (\ref{eq:D1}) becomes
\be
\De'' = \al{\De} \De' + \bt{\De} \De + \gm{\De} {\curly B} + \eta_{\De} 
{\curly K}\;,
\label{eq:D2}
\ee
where
\be
\al{\De} &=& -2 \frac{(1-3w^2)}{(1+w)} + \sfrac32 (1+w)\;, \nonumber \\
\bt{\De} &=& \frac32 \left[ 1+ \frac{w^2(5-3w)}{(1+w)} \right]
- \frac{8 (1-3w)^2}{9 (1+w)} k_E^2\;, \nonumber \\
\gm{\De} &=& -9w \caE \left[ \frac{1-w+2w^2}{(1+w)} \right]
-2 \caE (1-3w)^2 k_E^2\;, \nonumber \\
\eta_{\De} &=& 6w \caE \left[ \frac{(2+w+3w^2)}{(1+w)} \right]\;.  
\label{eq:delcoeffs}
\ee           
Similarly, (\ref{eq:B1}) and (\ref{eq:K1}) become
\be
{\curly B}' = \al B \De' + \bt B \De\;,
\label{eq:B2}
\ee
and 
\be
{\curly K}' = \al K \De + \bt K {\curly K} + \gm K {\curly B}\;,
\label{eq:K2}
\ee
where
\be
\al B &=& \frac{4}{3(1+w)}\;, \nn \\
\bt B &=& \frac{4w}{3 (1+w)^2} (1-3w)\;, \nn \\
\al K &=& \frac{16w}{3(1+w)^2}\;, \nn \\
\bt K &=& - \frac{8w}{(1+w)} \caE\;, \nn \\
\gm K &=& \frac{12w}{(1+w)} \caE\;,
\label{eq:BKcoeffs}
\ee 
with the evolution of the background determined by (\ref{eq:wprop}).
\subsection{Reducing the Order of the System}
In the simplest case of a single fluid, equations (\ref{eq:D2}), 
(\ref{eq:B2}) and (\ref{eq:K2}) together with the condition 
$w = {\rm constant}$, form a fourth\hs order autonomous system of
differential equations. In the more general two\hs component case, 
the system is five dimensional due to the inclusion of 
the propagation equation for $w$ (\ref{eq:wprop}). However, the order 
of the system can be reduced by introducing a new variable $\curly U$ 
\cite{BruPio}. This reduction makes it possible to easily determine
the qualitative behaviour of $\De$, rather than its exact evolution
law. We start by defining $X = \De'$ and
\be
{\curly U} = X / \De = \De' / \De\;, \nn \\
{\curly R} = \sqrt{\De^2 + X^2}\;.
\label{eq:Udef}
\ee
${\curly R}$ represents an `amplitude' of the
perturbation and it is not directly relevant to our analysis. 
In order to keep the system dimensionally consistent, we 
define two new variables, 
\be
V= \frac{{\curly B}}{\De}\;, \qu {\rm and } \qu
W = \frac{{\curly K}}{\De}\;.
\label{eq:BKdef}
\ee
In terms of ${\curly U}, V, W$, the equations (\ref{eq:D2}), 
(\ref{eq:B2}) and (\ref{eq:K2}) become
\be
{\curly U}' = - {\curly U}^2 + \al{\De} {\curly U} + \bt{\De}
+ \gm{\De} V + \eta_{\De} W\;,
\label{eq:Uprop}
\ee
\be
V' = \al B {\curly U} + \bt B - V{\curly U}\;,
\label{eq:Vprop}
\ee
\be
W' = \al K + \bt K W + \gm K V - W {\curly U}\;,
\label{eq:Wprop}
\ee
with the coefficients determined by (\ref{eq:delcoeffs}), 
\& (\ref{eq:BKcoeffs}), together with the propagation equation 
for $w$ (\ref{eq:wprop}).

\end{multicols}

\begin{table}
\begin{tabular}{|l|l|cccc||}  \hline
{\em Point} &  {\em Values $(w,U,V,W)$} &
\multicolumn{4}{c|}{\em Eigenvalues ($\lambda_{1,2,3,4}$)} \\ \hline  
{\bf R1} & {($\sfrac13, 0, - \sfrac{1}{\caE}, - \sfrac{1}{\caE}$)} 
&   {$\sfrac13$} & {$-4\caE$} & {$2$} & {$-1+2\caE$} \\ 
{\bf R2} & {($\sfrac13, -4\caE, 1, -\sfrac12 (\sfrac{1}{\caE} +3)$)} 
&   {$\sfrac13$} & {$4\caE$} & {$2 + 8\caE$} & {$-1 + 2\caE$} \\
{\bf R3} & {$\sfrac13, -1+2\caE, 1, -(1+7\caE)$}
&   {$\sfrac13$} & {$1-2\caE$} & {$1-6\caE$} & {$3 - 2\caE$} \\
{\bf R4} & {($\sfrac13, 2, 1, \sfrac12 (1+ 2\caE)$)} 
&    {$\sfrac13$} & {$-2$} & {$-2-4\caE$} & {$-3+2\caE$} \\ \hline 
\end{tabular}
\caption{Table of equilibrium points for $w=\sfrac13$}
\label{radtable}
\end{table}

\begin{multicols}{2}

\subsection{Analysis}
\label{Analysis}
The system is non\hs linear, however we can analyse it locally 
by linearising about any stationary points, without losing the details
of the qualitative dynamics \cite{Ulf,DSinC}.

From equation (\ref{eq:wprop}) we can see that stationary points exist 
for $w=\sfrac13$ and $w=0$. 

Since equation (\ref{eq:wprop}) decouples from the rest of the
system, the 3\hs dimensional subsystems corresponding to these values
of $w$ correspond to invariant sets describing the early radiation\hs 
dominated and late dust\hs dominated periods of dynamical evolution. 
We now consider these cases separately.
\subsection{Radiation Era: $w=\case{1}/{3}$} \label{sec:radiation}

On substituting $w=\sfrac13$ into (\ref{eq:delcoeffs}) and 
(\ref{eq:BKcoeffs}), the system for the radiation\hs dominated era 
becomes:

\be
{\curly U}' &=& -{\curly U}^2 + {\curly U} + 2 - 2\caE V + 4 \caE W\;, \nn \\
V' &=& {\curly U} - V{\curly U}\;, \nn \\
W' &=& 1 - 2\caE W + 3\caE V - W {\curly U}\;.
\label{eq:radsys}
\ee             
The equilibrium points for (\ref{eq:radsys}) can easily be determined
and are given in table \ref{radtable} above. 

The 3\hs dimensional space described by $\{{\curly U}, V, W\}$, 
and $w=\sfrac13$ (see figure I) is an invariant set, so we can
classify the equilibrium points according to the eigenvalues 
$\lambda_{2,3,4}$ (since $\lambda_1 =\sfrac13$ is not relevant). 

{\bf Point R1} is a saddle (not included in figure I 
because it is far away from the other points). Orbits close 
to it may initially evolve towards it, but end up evolving away 
again. At {\bf R1} ${\curly U} = 0$, so density perturbations 
neither grow or decay. Using equations (\ref{eq:Udef}) and 
(\ref{eq:BKdef}) it follows that $\Delta$, ${\curly B}$ 
and ${\curly K}$ are constant: 
\be
\De &=& const = C_0\;, \nn \\
{\curly B} &=& - \sfrac{1}{\caE} C_0\;, \nn \\
{\curly K} &=& - \sfrac{1}{\caE} C_0\;. \nn \\
\label{eq:R1sol}
\ee
{\bf Point R2} is also a saddle, but this time since ${\curly U}<0$, 
density inhomogeneities are decreasing. Solutions at {\bf R2} again 
follow from equations (\ref{eq:Udef}) and (\ref{eq:BKdef}):  
\be
\De &=& C_1 {(\sfrac{a}{a_E})}^{-4\caE}\;,  \nn \\
{\curly B} &=& \De = C_1 {(\sfrac{a}{a_E})}^{-4\caE}\;,   \nn \\
{\curly K} &\approx& -\sfrac{1}{2\caE} C_1 {(\sfrac{a}{a_E})}^{-4\caE}\;.
\label{eq:R2sol}
\ee
{\bf Point R3} is a node source, representing unstable equilibrium, 
so all orbits close to this point evolve away from it. Again since 
${\curly U} = -1 + 2 \caE<0$, solutions at {\bf R3} represent 
a decreasing density inhomogeneity:
\be
\De &=& C_2 {(\sfrac{a}{a_E})}^{(-1+2\caE)}\;,  \nn \\
{\curly B} &=& \De = C_2 {(\sfrac{a}{a_E})}^{(-1+2\caE)}\;,   \nn \\
{\curly K} &=& -(1+ 7 \caE)C_2 {(\sfrac{a}{a_E})}^{(-1+2\caE)}\;.  \nn \\
\label{eq:R3sol}
\ee
{\bf Point R4} is a stable node, or a sink. Orbits close to
this point evolve towards it as $(\frac{a}{a_E})$ increases. Since  
${\curly U}>0$ solutions at {\bf R4} represent growing density 
inhomogeneities:
\be
\De &=& C_3 {(\sfrac{a}{a_E})}^{2}\;,  \nn \\
{\curly B} &=& \De = C_3 {(\sfrac{a}{a_E})}^{2}\;,   \nn \\
{\curly K} &=& \sfrac12 (1+ 2 \caE)C_3 {(\sfrac{a}{a_E})}^{2}\;.  \nn \\
\label{eq:R4sol}
\ee
Solutions at the points given above correspond to approximate solutions 
(to leading order in $\caE$) of the perturbation equations during the 
radiation\hs dominated era. Therefore by linearity, the general
solutions for the perturbation variables $\Delta$, ${\curly B}$ 
and ${\curly K}$ are given by a linear combination 
of (\ref{eq:R1sol}\hs\ref{eq:R4sol}):
\be
\De &=& C_{0}+ C_1{(\sfrac{a}{a_E})}^{-4\caE} +
C_2{(\sfrac{a}{a_E})}^{(-1+2\caE)} + C_3 {(\sfrac{a}{a_E})}^{2} \nn \\
{\curly B} &=& - \sfrac{1}{\caE} C_0 
+ C_1{(\sfrac{a}{a_E})}^{-4\caE}\nn \\ 
&+& C_2{(\sfrac{a}{a_E})}^{(-1+2\caE)} 
+ C_3 {(\sfrac{a}{a_E})}^{2}\;, \nn \\
{\curly K} &\approx&  - \sfrac{1}{\caE}C_0 
- \sfrac{1}{2\caE} {(\sfrac{a}{a_E})}^{-4\caE}\nn \\
&-& (1+7 \caE)C_2{(\sfrac{a}{a_E})}^{(-1+2\caE)}
\nonumber \\ &+& \sfrac12 (1+2\caE) C_3 {(\sfrac{a}{a_E})}^{2}\;.
\label{eq:radsolutions}
\ee

This can be written more concisely as:

\end{multicols}

\begin{table}
\begin{tabular}{|l|l|cccc||}  \hline
{\em Point} &  {\em Values $(w,U,V,W)$} &
\multicolumn{4}{c|}{\em Eigenvalues ($\lambda_{1,2,3,4}$)} \\ \hline  
{\bf D1} & {$(0,0,\sfrac{1}{2  \caE} [\sfrac{3}{2 {k_E}^2} - \sfrac89]$, const)}
& {$-\sfrac13$} &  $0$ & $-\sfrac14 + \xi$ & {$-\sfrac14 - \xi$} \\
{\bf D2} & {$(0, [-\sfrac14 + \xi], \sfrac43, 0)$}
& {$-\sfrac13$} & {$-2 \xi$} & {$\sfrac14 - \xi$} & {$\sfrac14 - \xi$} \\
{\bf D3} & {($0, [-\sfrac14 - \xi], \sfrac43, 0$)}
& {$-\sfrac13$} & {$ 2\xi$} & {$\sfrac14 + \xi$} & {$\sfrac14 + \xi$} 
\\ \hline
\end{tabular}
\caption{Table of equilibrium points for $w=0$, with $\xi=\sfrac12
\sqrt{6(\frac{25}{24}-\frac{k_E^2}{k_{EC}^2})}$.}
\label{dusttable}
\end{table} 

\begin{multicols}{2}

\begin{figure}
\epsffile{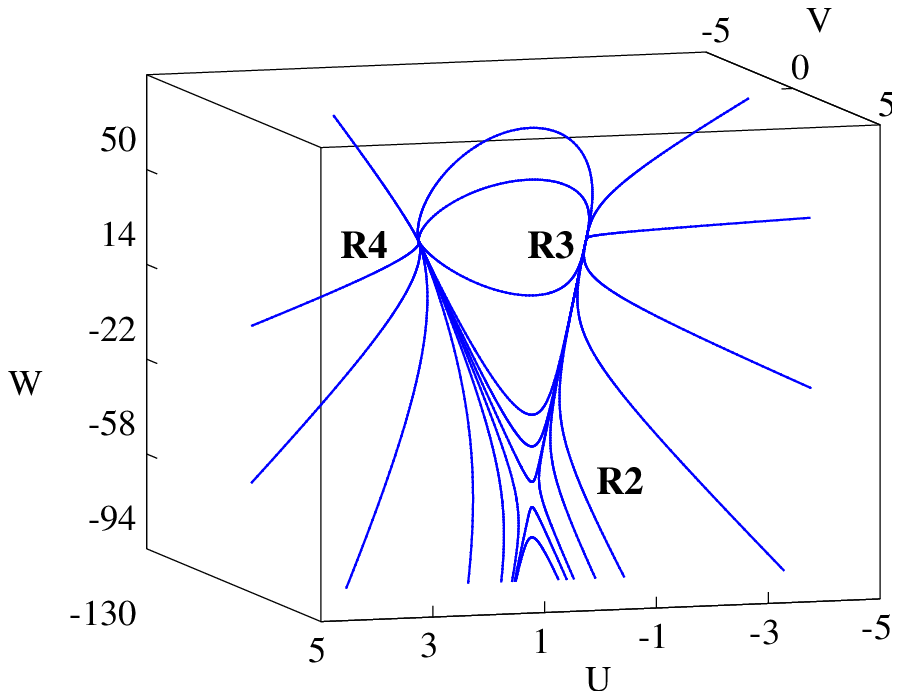}
\end{figure}
\begin{center}
Figure I: 
\end{center}

\be
\De &=& C_{(0)} + \sum_{\alpha} C_{(\alpha)} 
{(\sfrac{a}{a_E})}^{\alpha}\;, \nn \\
{\curly B} &=& -\sfrac{1}{\caE} C_0 + \sum_{\alpha} C_{(\alpha)} 
{(\sfrac{a}{a_E})}^{\alpha}\;, \nn \\
{\curly K} &=& - \sfrac{1}{\caE} 
+ (1 + 3\caE) \sum_{\alpha} \frac{C_{(\alpha)}}{(\alpha 
+ 2\caE)} {(\sfrac{a}{a_E})}^{\alpha}\;,
\label{eq:radconcise}
\ee
where $\alpha$ solves the cubic equation:
\be
{\alpha}^3 + (2\caE -1) \alpha^2 - 2 \alpha - 8 \caE (1+ \caE) = 0\;,
\ee
and corresponds to the super\hs horizon solutions given in \cite{TsaMar}, 
with slightly modified exponents \hs due to the extra term in the 
equation for the spatial curvature ${\curly K}$ (\ref{eq:K1}).
\subsection{Dust Era}
When solving for the equilibrium points in the dust\hs dominated 
era, we substitute $w = 0$ into the equations defining the 
dynamical system (\ref{eq:Uprop}\hs\ref{eq:Wprop}). 
This yields:
\be
{\curly U}'&=& -{\curly U}^2 - \sfrac12 {\curly U} + \sfrac32
- \sfrac89 {k_E}^2 - 2 \caE {k_E}^2 V\;, \nn \\
V'&=& \sfrac43 {\curly U} - V {\curly U}\;, \nn \\
W'&=& - W {\curly U}\;.
\label{eq:dustsys}
\ee
The equilibrium points of this system are shown in table \ref{dusttable}. 
In this case, the ${\curly U}$ and $V$ propagation equations decouple
from the equation for $W'$, which means that the system is effectively
only 2\hs dimensional. A critical scale $\lambda_{Ec}\equiv 2\pi
a/k_{Ec}$ appears in the analysis of these equation through its
corresponding wave number $k_{Ec}$:
\be
k^2_{Ec} = \frac{27}{16} \frac{1}{1+3\caE} \approx \frac{27}{16} 
(1 - 3\caE)\;. 
\ee
The solutions at each point will have different behaviour depending on
the value of $k_E$. In what follows, we first look at the general 
properties of the equilibrium points and their corresponding solutions 
and then give specific information about their behaviour for the three
regimes of $k_E$ values that emerge from the analysis.

{\bf Point D1:} These solutions are confined to the $\{{\curly U},V\}$
plane for constant $W$ and approach a constant value independent of
the value of $k_E$.
\be
\De &=& const  = C_0\;, \nn \\
{\curly B} &=& \sfrac{1}{2\caE} [\sfrac{3}{2 {k_E}^2} 
- \sfrac89] C_0\;, \nn \\
{\curly K} &=& const = C_k\;.  \nn \\
\ee

{\bf Point D2:} This point has different stability behaviour 
for different values of $k_E$. The curvature variable $W$ vanishes 
independently of $k_E$. Since the solution at the point depends on $\xi=\xi(k_E)$, the nature of the solution mode (e.g. growing/decaying/oscillation) 
will depend on the value of $k_E$:
\be
\De &=& C_1 ({\sfrac{a}{a_E}})^{(-\sfrac14 + \xi)}\;, \nn \\
{\curly B} &=& \sfrac43 C_1 ({\sfrac{a}{a_E}})^{(-\sfrac14 + \xi)}\;, \nn \\
{\curly K} &=& 0\;.  \nn \\
\ee
 
{\bf Point D3:} This point has different stability behaviour 
in two different regions of $k_E$. In both regions, the curvature 
variable vanishes. The solutions at {\bf D3} are:
\be
\De &=& C_2 ({\sfrac{a}{a_E}})^{(-\sfrac14 - \xi)}\;, \nn \\
{\curly B} &=& \sfrac43 C_2 ({\sfrac{a}{a_E}})^{(-\sfrac14 - \xi)}\;, \nn \\
{\curly K} &=& 0\;.  \nn \\
\ee
As in the radiation dominated era (discussed in section
\ref{sec:radiation}),  solutions (to leading order in $\caE$) for
the perturbation variables $\Delta$, ${\curly B}$ and ${\curly K}$ 
are given by a linear combination of the solutions at {\bf D1}, 
{\bf D2} and {\bf D3}:
\be
      \De &=&  C_0 + C_1 ({\sfrac{a}{a_E}})^{(-\sfrac14 + \xi)}
                    +  C_2 ({\sfrac{a}{a_E}})^{(-\sfrac14 - \xi)}\;, \nn \\
     {\curly B} &=& \sfrac{1}{2\caE} [\sfrac{3}{2 {k_E}^2} - \sfrac89] C_0  
               + \sfrac43 C_1 ({\sfrac{a}{a_E}})^{(-\sfrac14 + \xi)}
               +\sfrac43 C_2 ({\sfrac{a}{a_E}})^{(-\sfrac14 - \xi)}\;, \nn \\
     {\curly K} &=& const = C_k\;.  \nn \\
\ee
We now look more closely at how the nature of the stationary points 
{\bf D1}, {\bf D2} and {\bf D3} and their corresponding solutions 
depends on the wavenumber $k_E$. \\

\noindent \underline{{\it Region 1 - {$k_E \leq k_{Ec}$}}. See figure II}\\
    
{\bf Point D1:}
In this region, the eigenvalues are, respectively, $\lambda_2=0$, 
$\lambda_3\geq 0$, and $\lambda_4<0$. It follows that {\bf D1} corresponds 
to a line of saddles in the solution space for $\{U,V,W\}$.

{\bf Point D2:}
Here, all the eigenvalues are either negative or zero, so this point is 
a stable sink. Since $\xi \geq \sfrac14 \Rightarrow {\curly U} \geq 0$, 
it follows that the solution at {\bf D2} corresponds to a growing 
density inhomogeneity. 

{\bf Point D3:}
All eigenvalues are positive, so this point is an unstable
node. Since ${\curly U} = - \sfrac14 - \xi<0$,
it follows that the solution at {\bf D3} corresponds to a decaying 
density inhomogeneity. 

\begin{figure}
\epsffile{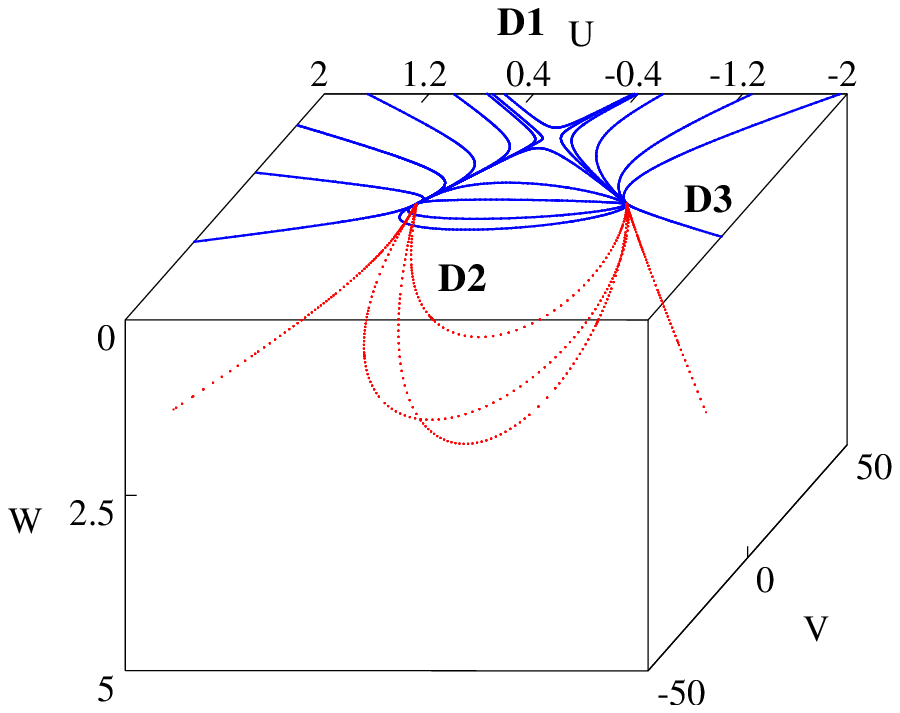}
\label{fig:dustsup}
\end{figure}
\begin{center}
Figure II: 
\end{center}
For very long wavelength solutions i.e. $k_E << k_{EC}$ we have 
$\xi \approx \sfrac54$. In this case the explicit solutions of 
the perturbation equations (which are a linear combination of the solutions 
at the points {\bf D1}, {\bf D2} and {\bf D3}) for the dust era are 
given by:
\be
\De &=& C_0 +  C_1 ({\sfrac{a}{a_E}})
+ C_2  ({\sfrac{a}{a_E}})^{-\sfrac32} \nn \\
{\curly B} &=& \sfrac{1}{2  \caE} [\sfrac{3}{2 {k_E}^2} - \sfrac89] C_0
+\sfrac43 C_1({\sfrac{a}{a_E}})
+ \sfrac43 C_2 ({\sfrac{a}{a_E}})^{-\sfrac32} \nn \\
{\curly K} &=& C_k\;.
\label{eq:dustsup}
\ee
In the dust era, $(\frac{a}{a_E}) = {(\frac{t}{t_E})}^{2/3}$, so we can write
the above solutions (\ref{eq:dustsup}) in terms of $(\frac{t}{t_E})$ as in 
\cite{TsaMar}. For example, the solution for density perturbations is: 
\be
\De &=& C_0 + C_1 {(\sfrac{t}{t_E})}^{2/3} + C_2
{(\sfrac{t}{t_e})}^{-1}\;.
\nn    
\ee
It consists of a magnetic field induced constant mode, plus the two 
usual non\hs magnetized adiabatic modes. Comparing to \cite{TsaMar} 
equation (65), we see that the dynamical systems analysis does not recover 
the non\hs adiabatic decaying mode. This is because our analysis is
set up to look for asymptotic solutions, so as $t\rightarrow \infty$, 
the magnetic energy density decays faster than that of the fluid, 
and $\ca \rightarrow 0$. This removes the magneto\hs curvature 
coupling from the propagation equation for ${\curly K}$, and therefore 
our analysis does not obtain the extra decaying mode induced by the 
magnetic field.\\

\noindent \underline{{\it Region 2 \hs  {$k_{Ec} < k_E \leq \frac{5}{2
        \sqrt 6} k_{Ec}$}}. See figure III}\\

{\bf Point D1:} In this region, the eigenvalues are, 
respectively, $\lambda_2=0$, $\lambda_3<0$ and $\lambda_4<0$. It 
follows that {\bf D1} represents a line of sinks (only one is shown in 
figure III) in the solution space for $\{U,V,W\}$.

{\bf Point D2:} In this region,  the eigenvalues are, 
respectively $\lambda_2\leq 0$, $\lambda_3>0$ and $\lambda_4>0$,
so {\bf D2} is a saddle point (or a line of sources if $\lambda_2=0$)
in the solution space for $\{U,V,W\}$. It follows, since ${\curly
  U}<0$, the solution at {\bf D2} represents a decaying inhomogeneity.

{\bf Point D3:}
Here we obtain the same behaviour as in the case when $k_E \leq
k_{EC}$, i.e. {\bf D3} is a saddle, corresponding to a decaying
density inhomogeneity.

\begin{figure}
\epsffile{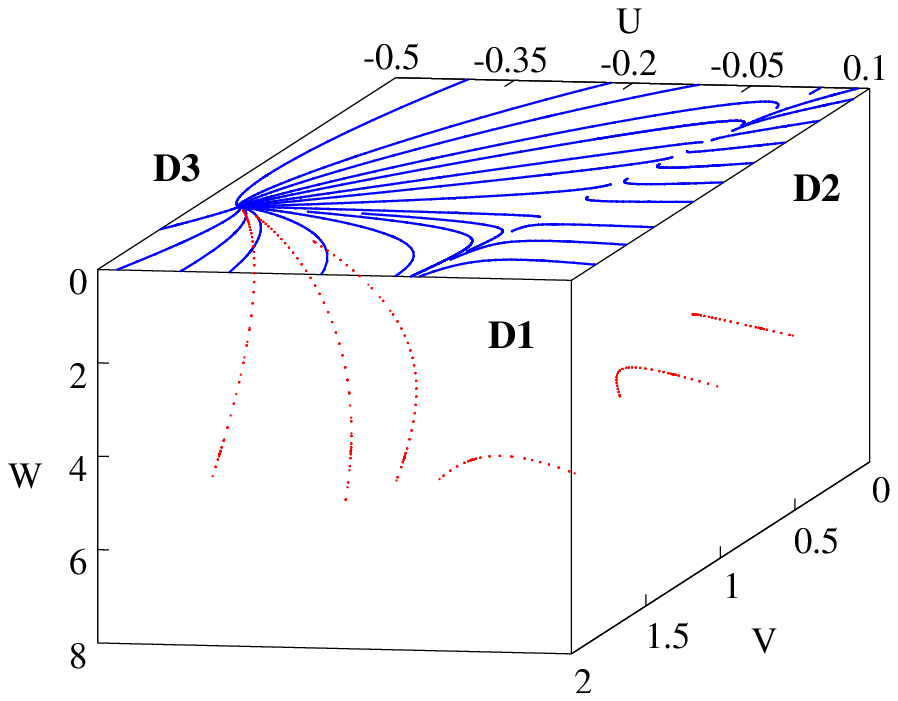}
\label{fig:dustmid}
\end{figure}
\begin{center}
Figure III: 
\end{center}

As for {\it region 1} we can now write down explicit solutions of 
the perturbation equations. In terms of proper time $\sfrac{t}{t_E}$ 
we obtain:
\be
      \De &=& C_0 + C_1 ({\sfrac{t}{t_E}})^{(-\sfrac16 + \xi_t)}
                    +  C_2 ({\sfrac{t}{t_E}})^{(-\sfrac16 - \xi_t)}\;, \nn \\
     {\curly B} &=& \sfrac{1}{2\caE} [\sfrac{3}{2 {k_E}^2} - \sfrac89] C_0  
               + \sfrac43 C_1 ({\sfrac{t}{t_E}})^{(-\sfrac16 + \xi_t)}
               +\sfrac43 C_2 ({\sfrac{t}{t_E}})^{(-\sfrac16 - \xi_t)}\;, \nn \\
     {\curly K} &=& const = C_k\;, \nn \\
\ee
where
\be
  \xi_t = \sfrac13 \sqrt{6(\frac{25}{24}-\frac{k_E^2}{k_{EC}^2})}
\ee
 and $0 \leq \xi_t < \sfrac16$.

In this region, none of the solutions correspond to growing modes 
for density inhomogeneities. The line of stable sinks  
represent modes where the density perturbation approaches 
constant value.\\

\noindent \underline{{\it Region 3 \hs
$k_E > \frac{5}{2 \sqrt 6} k_{Ec}$}. See figures IV and V}\\

{\bf Point D1:} For these values of $k_E$, the eigenvalues $\lambda_3$ and 
$\lambda_4$ are imaginary. Since their real parts are both 
negative, this solution acts as an attractor, and gives 
rise to a line of stable spiral points. This repetition of the
equilibrium points is, as mentioned before, a consequence of one of the
eigenvalues vanishing. Here, the density and magnetic parts of the 
solution decouple from the curvature ($W'$ equation), and thus the 
spiral point exists in any $W = {\rm const}$ plane (see figure V).

{\bf Point D2:} The solutions for ${\curly U}$ are imaginary in this region,
so density inhomogeneities oscillates as a sound wave, neither growing 
nor decaying.

{\bf Point D3:} Here there are no real solutions. Density
perturbations again oscillate as sound waves. 

This time the solutions are given by 
\be
      \De &=&  C_0 + C_1 ({\sfrac{a}{a_E}})^{(-\sfrac14 + {i \beta})}
                    +  C_2 ({\sfrac{a}{a_E}})^{(-\sfrac14 - {i \beta})}\;, \nn \\
     {\curly B} &=& \sfrac{1}{2\caE} [\sfrac{3}{2 {k_E}^2} - \sfrac89]
     C_0\nn \\  
               &+& \sfrac43 C_1 ({\sfrac{a}{a_E}})^{(-\sfrac14 + {i \beta})}
              +\sfrac43 C_2 ({\sfrac{a}{a_E}})^{(-\sfrac14 - {i \beta})}\;, \nn \\
     {\curly K} &=& C_k\;.
\ee
with $\beta=\sfrac12 \sqrt{6(\frac{25}{24}-\frac{k_E^2}{k_{EC}^2})}$.
In terms of the proper time $\sfrac{t}{t_E}$ we obtain:
\be
  \De &=& C_0 + (\sfrac{t}{t_E})^{-\sfrac16} 
        \bigl (C_3 \cos [\sfrac23 \beta \ln (\sfrac{t}{t_E})] + 
               C_4 \sin [\sfrac23 \beta \ln (\sfrac{t}{t_E})] \bigr )\;, \nn \\
{\curly B} &=&  \sfrac{1}{2\caE} [\sfrac{3}{2 {k_E}^2} -
\sfrac89]C_0 \nn \\
          &+& \sfrac43 (\sfrac{t}{t_e})^{-\sfrac16}
               \bigl (C_3 \cos [\sfrac23 \beta \ln (\sfrac{t}{t_E})] + 
               C_4 \sin [\sfrac23 \beta \ln (\sfrac{t}{t_E})] \bigr )\;,
               \nn \\
{\curly K} &=&  C_k\;.
\ee 
The spiral behaviour round the stable point is evident from
the damped oscillatory solution (see figures IV and V).

In this region, the wavelength of solutions falls below a critical
wavelength which is related to the Jeans length with a correction due
to the magnetic field (see section \ref{Jeans length}, below). 
Thus, these solutions do not result in growing density
inhomogeneities, but in general, oscillate as sound waves. The 
constant density solution due to the magnetic field 
(corresponding to {\bf D1}) represents the only stable solution 
in this region, and can be seen as a spiral point in figure V. Note 
that this is the only equilibrium point in this figure, as the 
other 2 solutions represent pure sound waves.

\begin{figure}
\epsffile{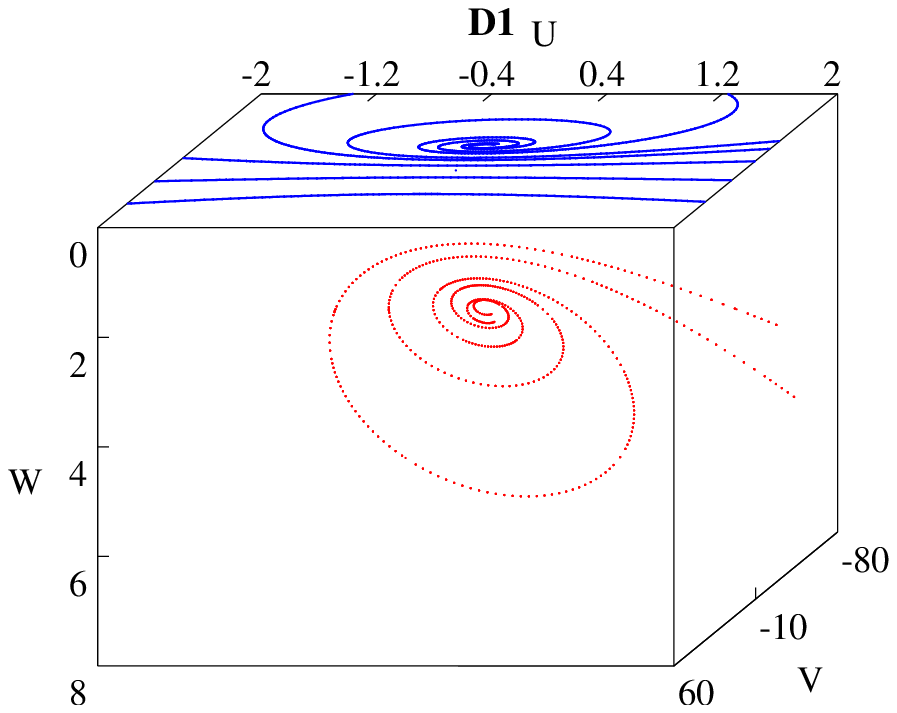}
\label{fig:smalldust}
\end{figure}
\begin{center}
Figure IV: 
\end{center}

\begin{figure}
\epsffile{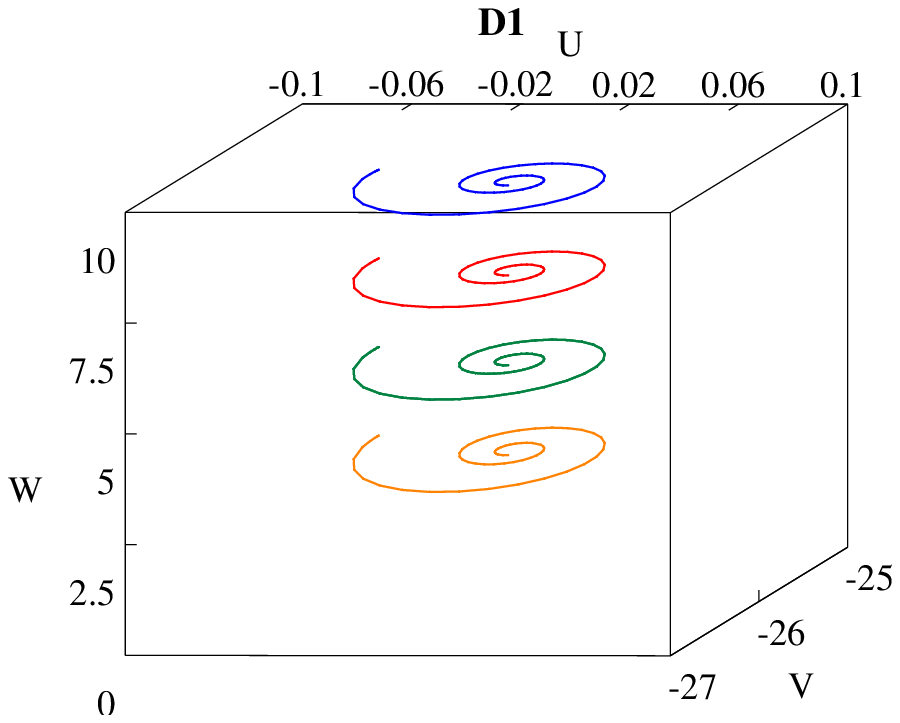}
\label{fig:spirals}
\end{figure}
\begin{center}
Figure V: 
\end{center}
\subsection{Magnetised Jeans length}
\label{Jeans length}
It is clears from the above discussions that the 
critical scale $k_{EC}$ which appears in our analysis is 
not quite the scale which determines the onset of oscillatory 
behaviour. This is determined by the wavenumber 
$k_{osc}=\sfrac{5}{2 \sqrt{6}} k_{EC}$, whose corresponding 
wavelength is:
\be
\lambda_{osc} = \sfrac{2 \sqrt{6}}{5} \lambda_{EC}
&=& \sfrac {2 \sqrt{6}}{5} 
[ \sfrac{2\pi}{H_E} \sfrac{4}{3 \sqrt{3}}(1+3 \caE)^{1/2}] \nn \\ 
&\approx& \sfrac{16\sqrt{2}\pi}{15} \sfrac{1}{H_E} (1 + \sfrac32 \caE)\;.
\label{eq:magJ}
\ee
This scale is closely related to the {\it Jeans Length} as it determines
the wavelength which divides oscillatory from growing or decaying 
solutions. It takes the form of a general critical wavelength,
modified by a linear factor due to the magnetic field. It is clear 
that since the magnetic field contributes to the overall pressure, 
it enlarges the size of regions able to resist gravitational collapse, 
and so increases the critical scale relative to the non\hs magnetised case.

In the limiting case $\caE \rightarrow 0$, we obtain:
\be 
\lambda_{osc} \rightarrow   \sfrac{16\sqrt{2}\pi}{15} 
\sfrac{1}{H_E}\;,
\ee
which is the corresponding scale for a flat dust\hs radiation model 
(with no magnetic field) found by Bruni and Piotrowska in \cite{BruPio}. 

The scale $\lambda_{osc}$ found above (\ref{eq:magJ}) 
differs considerably from that found in \cite{TsaMar} in their 
analysis of a magnetized dust solutions (see equation 68). In a 
pure dust model, 
the magnetic field acts as an effective pressure, allowing for 
the extraction of a `magnetic Jeans length`. This length is directly 
proportional to the Alfv{\'e}n speed, and vanishes if there is no 
magnetic field. Furthermore it is considerably smaller (of order $\caE$) 
compared to the critical length found by us using a two\hs fluid model, 
since in this case, the magnetic field is not the only source of 
pressure;  there is a contribution due to the radiation component.  
\section{Summary} \label{Summary}
Assuming a large\hs scale homogeneous magnetic field, we use the exact general 
relativistic propagation equations given in \cite{TsaBar14} to describe
the evolution of density and magnetic field inhomogeneities and curvature
perturbations. We use the usual approximations of small magnetic field
energy density ($B^2 / \rho << 1$), and infinite conductivity 
($\sigma \rightarrow \infty \Rightarrow E \rightarrow 0$), to 
simplify these equations. We carefully linearise the exact equations 
using a two\hs parameter approximation scheme and find an extra linear
term (compared to \cite{TsaBar14,TsaBar15}) in the propagation 
equation for the spatial curvature ${\curly K}$. This extra term
however makes no qualitative difference to the dynamics.

Rather than attempting to solve these equations analytically, we
follow Bruni and Piotrkowksa \cite{BruPio} by setting up the linear
perturbation equations as a four\hs  dimensional autonomous dynamical
system. This approach provides an elegant description of the dynamics 
of gravitational and magnetic inhomogeneities for a two component medium
comprising of pressure\hs free matter (dust) and radiation, 
interacting only through gravity.  
The equilibrium points for the three\hs dimensional invariant sets 
representing the radiation ($w=\case{1}/{3}$) and dust ($w=0$) eras
correspond to the approximate solutions of the perturbation equations  
derived in Tsagas and Maartens \cite{TsaMar}. For example in the 
radiation era we find the following solution for density perturbations 
on scales much larger than the Hubble radius:
\be
\De &=& C_{0}+ C_1{(\sfrac{a}{a_E})}^{-4\caE} + C_2{(\sfrac{a}{a_E})}^{(-1+2\caE)}+ C_3 {(\sfrac{a}{a_E})}^{2}\;.
\ee
The adiabatic decaying mode decays even less rapidly due to the 
magnetic field, and the decaying non\hs adiabatic mode decays slightly 
faster than the corresponding solution in \cite{TsaMar}. To zeroth
order in $\caE$, these solutions correspond exactly to those given 
in \cite{TsaMar} (see equations (50) and (54) in their paper). 

The extra term found in the linearisation of the curvature 
propagation equation is proportional to $\caE$, and thus tends  
asymptotically to zero. Since the dynamical systems analysis examines 
the asymptotic behaviour of these models, this extra term makes no 
difference in the analysis of the dust era.  Similarly, the superhorion
dust solutions do not include the non\hs adiabatic decaying mode 
induced  by the magnetic field, since the Alfv{\'e}n speed 
({$\ca=B^2/ \rho$}) also tends to zero at late times.

An important feature that arises out of the dynamical systems
analysis, is that we obtain three distinct evolution regimes for the 
perturbation modes. These regimes are defined through a critical scale
$\lambda_{EC}$ which is of the order of the Hubble scale at matter\hs
radiation equidensity and includes a modification linear in
$\caE$ due to the presence of the background magnetic field:
\be
\lambda_{EC} = \sfrac{1}{H_E} \sfrac{8\pi}{3 \sqrt{3}}(1+3 \caE)^{1/2}\;. \nn
\ee
The three regimes are:
\begin{itemize}
\item{Large wavelengths \hs  $\lambda_E > \lambda_{EC}$}
These are large\hs scale density and magnetic inhomogeneities that
grow unbounded giving gravitational instability. 
Solutions for these wavelengths consist of a growing mode, 
a decaying mode, and a constant mode, represented by a saddle point 
in the phase space of solutions.
\item{Intermediate wavelengths \hs $\lambda_{osc} < \lambda_E < \lambda_{EC}$}
These perturbations are over\hs damped and therefore decay asymptotically
to a constant value $C_0$ without oscillating.
\item{Small wavelengths \hs $\lambda_E < \lambda_{osc}$}
Here the perturbations oscillate like sound waves, while their 
amplitude decays.
\end{itemize}

Furthermore, the analysis of the dust models yields a magnetic field
corrected scale $\lambda_{osc}$ closely related to the 
{\it Jeans length}, which is a multiple of the general critical 
scale $\lambda_{EC}$ for flat radiation\hs dust  models:
\be
\lambda_{osc} =  &=& \sfrac {2 \sqrt{6}}{5} 
[ \sfrac{1}{H_E} \sfrac{8\pi}{3 \sqrt{3}}(1+3 \caE)^{1/2}] \nn \\ 
&\approx& \sfrac{16\sqrt{2}\pi}{15} \sfrac{1}{H_E}
(1 + \sfrac32 \caE)\;,
\ee
where the quantities {$\caE,~H_E,~\lambda_E$} are evaluated at equi\hs density.

Finally we note that $\lambda_{osc}$ is more general 
than the `magnetised Jeans Length` found in  \cite{TsaMar} for a 
pure dust model, since it takes into account the pressure effects 
resulting from a proper two\hs fluid description.
\section*{Acknowledgments}
We thank Mattias Marklund, Christos Tsagas and Roy Maartens for
helpful discussions, the referee for his/her comments and the 
NRF (South Africa) for financial support.
\begin{appendix}
\section{Covariant identities} \label{ids}
This identities are written using the notation of \cite{maartens} 
and \cite{TsaMar}, except using $\tnabla$ instead 
of $D$ for the orthogonally projected covariant derivative. 
These identities are used in deriving the propagation equations 
(assuming a flat background and vanishing cosmological constant):
\be
{\rm curl} \tnabla_a f &=& -2 \dot{f} \omega_a\;, \\
{(a \tnabla_a f)}^. &=& a \tnabla_a \dot{f} + a \dot{f} A_a\;,\\
\tnabla^2(\tnabla_a f) &=& \tnabla_a (\tnabla^2 f) 
+ 2 \dot{f} {\rm curl} \omega_a\;, \\
(a \tnabla_a J_b..)^. &=& a \tnabla_a {\dot{J}}_b...\;, \\
\label{eq:magdif}
\tnabla_{[a} \tnabla_{b]} V_c &=& 0 = \tnabla_{[a} \tnabla_{b]} S^{cd}\;, \\
{\rm div}~{\rm curl~} V &=& 0\;,\\
({\rm div }~{\rm curl~} S)_a &=& \sfrac12 {\rm curl}({\rm div} S)_a\;,\\
{\rm curl~curl~} V_a &=& \tnabla_a ({\rm div} V) - \tnabla^2 V_a \\
{\rm curl~curl~} S_{ab} &=& \sfrac32 \tnabla_{\la a} ({\rm div} S)_{b \ra} 
- \tnabla^2 S_{ab}\;,
\ee
where the vectors and tensors vanish in the background 
and $S_{ab} = S_{\la ab \ra}$. The magnetic field vector does 
not vanish in the background, and so its projected derivatives 
do not commute to linear order. For the magnetic field, the vector
identity in (\ref{eq:magdif}) must therefore be changed to:
\be
\tnabla_{[a} \tnabla_{b]} B_c = \sfrac12 {\curly R}_{dcba} B^d                 -{\veps}_{abd} \omega^d {\dot{B}}_c\;,
\ee
where ${\curly R}_{abcd}$ is the 3\hs Curvature tensor 
formed from $R_{abcd}$ and the kinematic quantities 
\cite{TsaBar14,TsaBar15}.

\section{Harmonic Decomposition} \label{sec:appendB}
When setting up and analysing the dynamical system, all the equations 
can be reduced to ordinary differential equations if we restrict our 
attention to the harmonic components of the perturbation
variables. This is a way of effectively separating the time from the 
space variables, and involves writing the perturbation scalars in terms
of harmonic scalars as follows \cite{Tsagas} :
\be
\De &=& \sum_{n} \De_{(n)} Q_{(n)}\;,  \nn \\
{\curly B} &=& \sum_{n} {\curly B}_{(n)} Q_{(n)}\;, \nn \\
{\curly K} &=& \sum_{n} {\curly K}_{(n)} Q_{(n)}\;. \\
\label{harmonics}
\ee
The scalar harmonics, $Q$, are defined by 
\begin{eqnarray}
\dot{Q_{(n)}} &=& 0 \qu {\rm and} \nn \\
\tnabla^2 Q_{(n)} &=& - \frac{n^2}{a^2} Q_{(n)}\;, \\
\label{eq:harm}
\end{eqnarray}
where $n= k \geq 0$ (since we are dealing with a flat model). 
Spatial flatness of the background also means that $k$ is simply 
related to the wavelength $\lambda$ of the perturbation since 
$\lambda = \frac{2 \pi a}{k}$.
\section{Coefficients in terms of $w$}  \label{sec:appendC}
$\cs$ is only formally the speed of sound since the fluids interact 
only through gravity. Using the energy\hs density conservation
equations for matter and radiation we can express $\cs$ in terms of
the parameter $w$: 
\be
\cs = \frac{\dot{p}}{\dot{\rho}}
=  \frac{4w}{3(1+w)}\;.
\label{eq:cs}
\ee
The total energy density of the matter\hs radiation mixture 
is given by:
\be
\rho = \frac{27}{2} \rho_E \frac{w^3}{(1-3w)^4}\;,
\ee
where $\rho_E$ is the energy density at matter\hs radiation equality.
This leads to a similar equation for the Hubble parameter $H^2$:
\be
H^2= \frac{27}{2} H^2_E \frac{w^3}{(1-3w)^4}\;,
\ee
where $H_E$ is the value of the Hubble parameter at matter\hs
radiation equality.

When harmonically decomposing the perturbation equations in section \ref{DS},
the Laplacian terms give rise to coefficients of the form  $k^2/a^2
H^2$. These can be expressed in terms of the variable $w$ as follows:
\be
\frac{k^2}{a^2 H^2} = \frac{k^2}{a_E^2 H_E^2} \frac{2(1-3w)^2}{3w} 
\nonumber \\
= k_E^2. \frac{2(1-3w)^2}{3w}\;,
\ee
where $k_E^2$=$\frac{k^2}{a_E^2 H_E^2}$.
\end{appendix}

\end{multicols}

\end{document}